\documentclass[runningheads,fleqn]{svmult}

\usepackage{makeidx}   
\usepackage{graphicx}  
\usepackage{subeqnar}  
\usepackage{multicol}  
\usepackage{physmult}  
\makeindex             

\begin{document}
\title*{Patterns, Trends and Predictions \\in stock market
indices \\and foreign currency exchange rates}
\toctitle{Patterns, Trends and Predictions \protect\newline
in stock market
indices \protect\newline and foreign currency exchange rates}
%
%
\titlerunning{Patterns, Trends and Predictions in stock market}
\author{ Marcel Ausloos\inst{1} \and Kristinka Ivanova\inst{2}}

\authorrunning{M. Ausloos and K. Ivanova}

\institute{ SUPRAS \& GRASP, B5, University of Li$\grave e$ge, B-4000 
Li$\grave e$ge, Belgium \and Pennsylvania State University, 
University Park PA 16802, USA}

\maketitle
\begin{abstract}
Specialized topics on financial data analysis from a numerical and 
physical point
of view are discussed. They pertain to the analysis of crash 
prediction in stock
market indices and to the persistence or not of coherent and random 
sequences in
fluctuations  of foreign exchange currency rates. A brief historical 
introduction
to crashes is given, including recent observations on the DJIA and the S\&P500.
Daily data of the DAX index are specifically used for illustration. The method
for visualizing the pattern thought to be the precursor signature of financial
crashes is outlined. The log-periodicity of the pattern is investigated.
Comparison of patterns before and after crash days is made through the power
spectrum.  The corresponding fractal dimension of the signal looks 
like that of a
percolation backbone. Next the fluctuations of exchange rates (XR) of 
currencies
forming $EUR$ with respect to $USD$ are analyzed.  The XR power spectra are
calculated before and after crashes. A detrended fluctuation analysis is
performed. The characteristic exponents $\beta$ and $\alpha$ respectively, are
compared, including the time dependence of each $\alpha$, found to be singular
near crash dates.
\end{abstract}

\section{An Introduction with Some Historical Notes as "Symptoms"}

The stock market crash on Monday Oct. 19, 1987 led to the {\it October black
monday syndrome}. On that day, the Dow Jones Industrial Average (DJIA) lost
21.6~\%. Other markets were shaken : the worst decline reached 45.8~\% in Hong
Kong. The downturn was spread out over two or three days in different European
stock markets: the DAX lost 10~\%. Nevertheless most markets had been 
using for a
long time {\it breakers}, i.e. periods of trading halts and/or limitations of
daily variations. This tends to suggest that the adoption of circuit 
breakers at
the very least do delay the crash process, and not much more.  In fact,
Lauterbach and Ben-Zion \cite{lb1993} found that trading halts and price limits
had no impact on the overall decline of October 1987, but merely 
smoothed return
fluctuations in the neighborhood of the crash.

Another major characteristic of the crash of October 1987 is the phenomenon of
irresistible contagion. It is well accepted that the shock arose 
first from Asian
markets, except Japan, then propagated to the European markets before reaching
the American markets after, when Asian markets were already closed
\cite{Roll1989}. A mapping of the Nikkei, DAX and DJIA daily sign fluctuations
has been made onto a 1/2 Ising spin chain as if there was a continuous index
calculated three times during 24 hours. This showed that the spin cluster
fluctuations are rather equivalent to random fluctuations, - except during
pre-crash periods in which ($ down-down-down$) spin clusters form with a higher
probability than expected if the fluctuations are to be considered 
independent of
each other \cite{domino}. This has allowed to eliminate all 
criticisms about the
major responsibility of the derivative markets in the United States on the Oct.
87 crash. Since then, the world-wide interdependence of the economy has been
going on still more strongly. Thus if really a speculative bubble is 
occurring on
some financial markets, as commonly observed in the recent years, the 
phenomenon
of propagation to other stock exchanges could be  more important now. 
It is known
that methods of negotiation have widely changed on many financial 
places. Markets
using electronic systems of negotiation take advantage of recent improvement in
their transaction capacity. It is easier today to face a substantial 
increase in
transaction volumes during a major crisis period. Moreover, the 
efficient use of
the derivative markets could avoid useless pressures on the traditional market.
These financial factual observations should be turned into 
quantitative measures,
in order to, if necessary, avoid crashes.

Whence there is a need for techniques capable of rapidly following a bubble
explosion or preventing it. Notice that the drop of stock market 
indices can not
only spread out over two or three days but also over a much longer period. The
example of the Tokyo stock exchange at the beginning of the 1990's is 
a prominent
illustration. By comparison to the most famous crash of 1929, the Oct. 87 crash
was spread over 2 days: the Dow Jones sank 12.8~\% on October 28 and 11.7~\% on
the following day. (That was similar for the DAX which dropped by 8.00~\%  and
7.00~\% on Oct. 26 and 28, 1987 respectively.) This shows that a stock market
index decline does not necessarily lead to a crash in one single day. 
Indeed, the
decline can be slow and last several days or even several months in 
what would be
called not a crash, but a long duration bear market.

In the present econophysics research context, it is of interest to examine
whether the evolution of quotations on the main stock exchange places have
similarities and whether crash symptoms can be found. Even if history generally
tends to repeat, does it always do so in similar ways, and what are the
differences? A rise in quotations can be interpreted {\it a posteriori} as the
result of a speculative bubble but could be mere euphoria. How this 
does lead to
a rupture of the trend? Can the duration differences be interpreted? 
Can we find
universality classes?

Physics-like model of fracture or other phase transitions, including 
percolation
can be turned into some economic advantage. Along the same lines of 
thought, the
question was already touched upon in \cite{bigtokyo} within the sand 
pile model.
This allows not only a verbal analogy of index rupture in terms of sand
avalanches, but also some insight into the mechanisms. Through 
physical modeling
and an understanding of parameters controlling  the output, as in the sand pile
model, symptoms can be measured, whence to suggest remedies is not impossible.

Another question raised below is the post-crash period. One might expect from a
physics point of view that if a crash looks like a phase transition, and is
characterized by scaling laws, as we will see it sometimes occurs
\cite{SJB96,phasetr},  it might be expected that a relation exists between
amplitudes and laws on both sides of the crash day \cite{roehner4}. 
As mentioned
above, the crash might be occurring on various days, with different breaks. It
might be possible that between drops some positive surge might be found. Thus
some sorting of behaviors into classes should be made  as well. {\it In fine},
some discussion on the foreign exchange market will be given in order to recall
the detrended fluctuation analysis method, so often used nowadays. It 
is applied
below to the $USD$ exchange rate vs. the (ten) currencies forming the $EUR$ on
Jan. 01, 2000, over a time interval including the most recent years. The
observation of the time variation of the power law scaling exponent of the DFA
function is shown to be correlated to crash time occurrence.

\subsection{Tulipomania}

In  1559, the first tulip bulb (TB) was brought to Holland from China by Conrad
Guenster \cite{webtulip}. In 1611, the tulip bulbs (TBs) were stocked 
and sold on
markets. In 1625, one tulip bulb was worth 5 dutch gulden (NLG). The flower was
considered so rare that wealthy aristocrats and merchants tripped 
over themselves
to buy the onions. Speculation ensued and the TBs became wildly overvalued. The
TBs were not necessarily planted, but were just stored in the house salon. In
1635, 1 TB was worth 4 tons of wheat + 4 oxen  + 8 tons rye + 8 pigs 
+ one bed +
12 sheep + clothes + 2 wine casks + 4 tons beer + 2 tons butter + 1000 pounds
cheese + 1 silver drinking cup. In 1637, 1 TB was worth 550 NLG. One average
house was worth 17 800 NLG, whence about 30 TBs. However within 1637, over a 6
week time span the price of 1 TB went down 90~\%.

In view of the shock, remedies  had to be found and people called upon the
Amsterdam Parliament  for legislation. It was decided that all 
contracts would be
void if they were dealt before Nov. 1636, and after that date the 
contracts kept
a 10~\% value. Under some protest, people appealed to the Netherland Supreme
Court which ruled that this business of selling/buying TBs was mere 
gambling, and
no debt could be defined "by law"  nor ruled upon. Nowadays one TB is worth 0.5
EUR. Too bad for long term investment strategies. The TB became the classical
example for illustrating the {\it Extraordinary Popular Delusions and 
the Madness
of Crowds} as described by Charles MacKay \cite{webmackay}.

Just for the sake of physics history, let it be recalled that 1639 was 
the year in which Galileo Galilei (Pisa, Feb. 15, 1564; Arcetri,  Jan. 8, 1642) betrayed
science in saving his life.

\subsection{Monopolymania}

Another set of financial crises is that of the Compagnie du Mississipi
\cite{weblaw}  and that of the South Sea Company
\cite{websouthsea1,websouthsea2}. In 1715, John Law, a scot gambler, had
persuaded Philippe, Regent of France, to consider a banking scheme 
that promised
to improve the financial condition of the kingdom. In theory a private affair,
the system was linked from the beginning with liquidating the 
national debt. When
the monopoly of the Louisiana trade was surrendered in 1717, Law created a
trading company known as the {\it Compagnie d'Occident} (or {\it Compagnie du
Mississipi}) linked to the Royal Bank of France (first chartered in 1716 as
Banque G\'en\'erale) and in which government bills were accepted for the
purchase of shares.

Law gained a monopoly on all French overseas trade. The result was a 
huge wave of
speculation as the value of a share went from its initial value, i.e. 500 {\it
livres} to 18 000 $livres$. When the paper money was presented at the bank in
exchange for gold, which was unavailable, panic ensued, and shares felt by a
factor of 2 in a matter of days.

In England, the Whigs  represented the mercantile interests which had profited
from the War of the Spanish Succession War (1703-1711), and made 
large profits by
financing it, in doing so had created a National Debt which had to be 
financed by
further taxation.  During the wars the government handled more money than ever
before in history, and they skimmed off a lot through various 
methods, including
the invention of the Bank of England in 1694. The South Sea Company 
was formed in
1711 by the Tory government of Harley to trade with Spanish America, and to
offset the financial support which the Bank of England had provided 
for previous
Whig governments. They had in mind to establish a system like the Compagnie du
Mississipi Monopoly, \cite{weblaw} using the same sort of trading 
privileges and
monopolies, those granted to Britain after the Treaty of Utrecht. King George I
of Great Britain became governor of the company in 1718, creating confidence in
the enterprise, which was soon paying 100 percent interest. In 1720 a bill was
passed enabling persons to whom the government owed portions of the 
national debt
to exchange their claims for shares in company $stocks$, and to 
become {\it stock
holders}. In the 1719-20 the England Public Debt went to South Sea 
Company stock
holders, as approved by Parliament. On March 1 the stocks were valued GBP 175,
moved quickly to 200. Shortly the directors of the South Sea Company 
had assumed
three-fifths of the national debt.  The company expected to recoup itself from
expanding trade, but chiefly from the foreseen rise in the value of its shares.
On June 1 the shares were valued 500 and more than 1,000 in August 1720. Those
unable to buy South Sea Company stocks were inveigled by overly optimistic
company promoters or downright swindlers into unwise investments. Speculators
took advantage of investors to obtain subscriptions for sensibly unrealistic
projects. By September 1720 however, the market had collapsed, and by December
1720 South Sea Company shares were down to 124, dragging others, including
government stocks with them. Many investors were ruined, and the 
House of Commons
ordered an inquiry, which showed that  ministers had accepted bribes and
speculated. From a physics point of view let it be recalled that I. Newton
(Woolsthorpe, Dec. 25, 1642; London, March 20, 1727) invested in such South Sea
Company stocks and lost quite a bit of money \cite{Newton}.

\subsection{WallStreetmania}

In the years from 1925 to 1929  one could easily go to a broker and purchase
stocks on margin, i.e. instead of buying stocks with real cash money, one could
purchase them with some cash down and the rest on credit. The Coolidge
administration had a {\it laissez-faire policy}, i.e. a government policy of
non-intervention. It was almost {\it la fa\c{c}on de vivre} to play 
in the stock
market \cite{wallstreetcrash29}. That allowed a speculation bubble to grow
unchecked. The Federal Reserve powers on economic matters were not utilized as
could be done nowadays. Many successions of $mini$ crashes and rallies began as
early as March 1929. The summer of 1929 hearkened somewhat of the good old days
of optimism. The market appeared to be stable. On Sept. 3, a bear market became
firmly established, and on Thursday Oct. 24, 1929 the famous 1929 
crash occurred
(Fig. 1).

\begin{figure} \includegraphics[width=.9\textwidth]{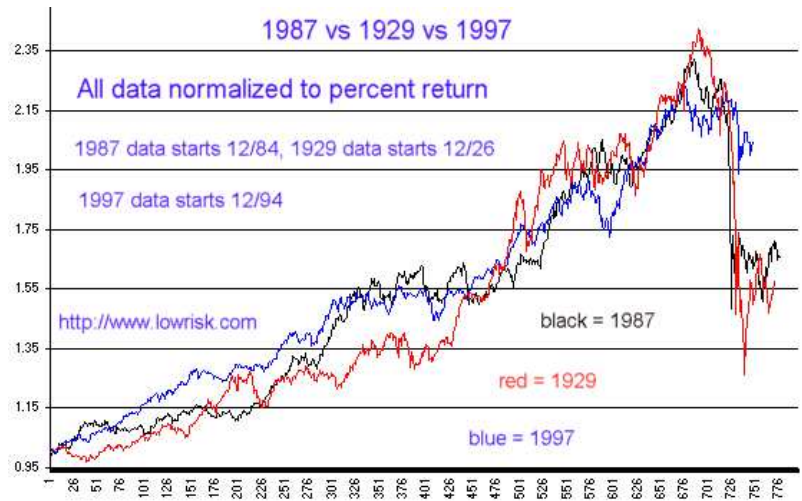} 
\caption[]{Famous
Wall Street Crashes : 1929, 1987, 1997; taken from color web site
  http://www.lowrisk.com/crash/87vs97.htm; from top to bottom on
day 201 : blue (97), red (29), black (87); on day 700 : red (29), black(87),
blue(97)} \label{eps1} \end{figure}

The 1987, 1997, and  more recent  1999, 2000, 2001 crashes are reminiscent and
even copies of the above ones (Fig. 1). The symptoms look similar : 
artificially
built euphoria, malingnantly established speculation, easy access to market
activities, including manipulated (or rather $electronipulated$) 
informations ...
Consider the buying frenzy on IPOs stocks at the end of the 1990's in 
companies
for which owners do not have a coherent business plan, or are not going to make
money, ... and yet see how we bought e-stocks. Nothing has changed since 1600,
1700 nor 1929.

\section{Econophysics of Stock Market Indices}

Econophysics \cite{ausloos,contwille} aims to fill the huge gap separating
''empirical finance'' and "econometric theories". Various subjects have been
approached like the option pricing, stock market data analysis, 
market modelling
and forecasting, etc...The application of statistical physics ideas to the
forecasting of stock market behavior has been proposed earlier following the
pioneer work of physicists interested by economy laws
\cite{mandel,stanley,bouch,peters1,peters2,MantegnaStanleybook,voit}.

Even though a stock market crash is considered as a highly unpredictable event,
it should be reemphasized that it takes place systematically during a period of
generalized anxiety spreading over the markets following a euphoria time. The
crash can be seen as a natural correction bringing the market to a "normal
state". Three important facts should be underlined:

(i) The series of daily fluctuations, so called $volatility$,  of the stock
market presents a huge clustering around the crash date, i.e. huge fluctuations
are grouped around the crash date. This is well illustrated in Fig. 2 for the
case of the DAX around 1987. The time span of this clustering is quite long: a
few years. This clustering indicates that larger and larger fluctuations take
place before crashes.

\begin{figure}
\includegraphics[width=.9\textwidth]{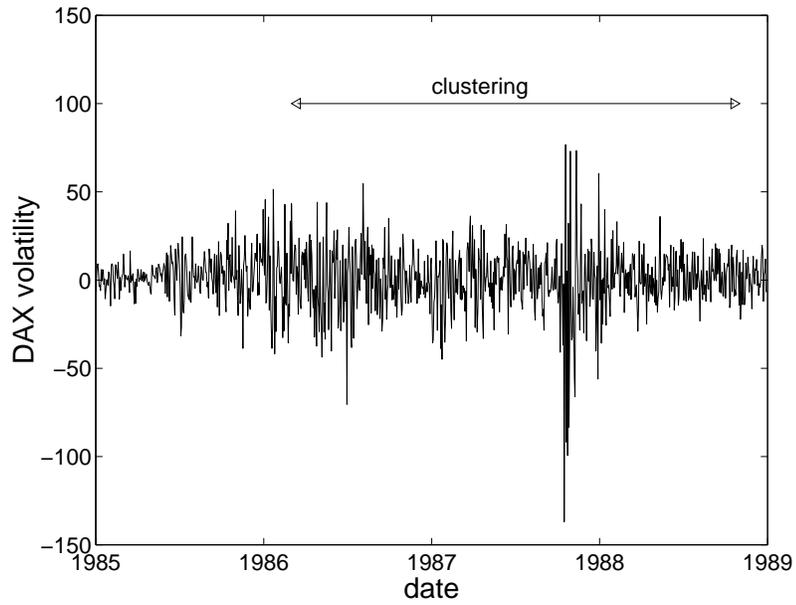}
\caption[]{DAX volatility between Jan. 01, 85 and Dec. 31, 88}
\label{eps2}
\end{figure}

(ii) Collective effects are to be considered, be they stemming from 
macroeconomy
informations, as a set of "external fields", and leading to a bear market, or
more intrinsically  $self-organized$, as if microeconomic informations  (or
$interactions ?$) were triggering the non-equilibrium state evolution.

(iii) A third remark concerns the panic--correlations appearing before crashes.
This kind of collective behavior is commonly observed during a trading day. The
market in Tokyo closes before London opens and thereafter New York 
opens. During
periods of panic, financial analysts are looking for the results and 
evolution of
the geographically preceding market. Strong correlations are found  in
fluctuations of different market indices before crashes.

\begin{figure}
\includegraphics[width=.9\textwidth]{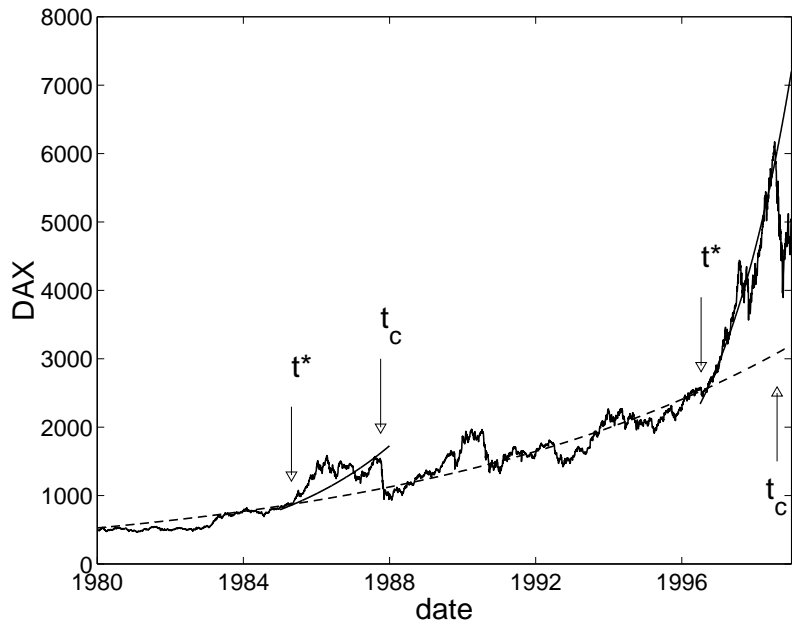}
\caption[]{DAX evolution between Jan.01, 1980 and Dec.31, 1998 with the mean
field behavior, the time(s) $t^*$, corresponding to the $Ginzburg-Levanyuk$
temperature bounding the critical fluctuation region, and the 
critical crash day(s)
$t_c$ }
\label{eps3}
\end{figure}

Of course, fluctuations and correlations are both ingredients which are
supposedly known to play an important role in phase transitions. Thus 
an analogy
can be derived (Fig.3) between phase transitions and crashes, defining the mean
field (exponential-like) behavior, the time $t^*$, corresponding to the
$Ginzburg-Levanyuk$ temperature bounding the critical fluctuation region, the
critical crash day $t_c$, etc. \cite{phasetr}. The character of a thermodynamic
phase transition is characterized by critical exponents, following the scaling
law hypothesis, exponents which are thought to depend on the symmetry of the
order parameter and the underlying lattice dimensionality\cite{StanleyPTbook}.
Similar considerations are looked for in financial crash studies.

In 1996, two independent works \cite{sornette,FF96} have proposed that critical
phenomena would be possible scenarios for describing crashes.  Those 
authors are
still debating about the subject \cite{sornette2,feig2}. More precisely, it has
been proposed that an economic index $y(t)$ increases as a power law decorated
with a log-periodic oscillation, i.e. \begin{equation} y = A + B {\left( {t_c-t
\over t_c} \right)}^{-m} \left[ 1 + C  sin\left( \omega \ln{\left( {t_c-t \over
t_c} \right)} + \phi \right) \right] \hskip 0.3cm  for  \hskip 0.3cm t < t_c
\end{equation} where $t_c$ is the crash-time or rupture point, $A$, $B$, $m$,
$C$, $\omega$ and $\phi$ are free parameters. This evolution $y(t)$ is in fact
the real part $\Re$ of a power law behavior at $t=t_c$ with a complex 
exponent $m
+ i \omega$, i.e. \begin{equation} y \sim \Re \left \{ {\left( t_c - t \over
t_c\right)}^{-m+i\omega} \right \} \end{equation}The law for $y(t)$ diverges at
$t=t_c$ if $m > 0$. This evolution is decorated with oscillations converging at
the rupture point $t_c$. This law is similar to that of critical points, and
generalizes the situation for cases in which a hierarchical lattice structure
exists, in other words  a Discrete Scale Invariance (DSI) is subjacent
\cite{dsi}.

The relationship (1) has been proposed elsewhere in order to fit experimental
measurements of sound wave rate emissions prior to the rupture of heterogeneous
composite stressed up to failure \cite{fracture}. The same type of 
complex power
law behavior has been also observed as a precursor of the Kobe earthquake in
Japan \cite{kobe}. Such log-periodic corrections have been recently reported in
biased diffusion on random lattices \cite{bias}.

As early as April 1997, Vandewalle and Ausloos performed a series of
investigations in order to emphasize crash precursors 
\cite{dup1,cash1,vif}. The
closing values of the Dow Jones Industrial Average (DJIA) and the Standard \&
Poor 500 (S\&P500) were used for tests. A law slightly different from 
Eq.(1) was
proposed \cite{dup1,how}. A strong indication of a so-called crash event or
market rupture point was numerically discovered \cite{dup1,cash1,vif,how}.
Further data analysis (in Aug. 97) \cite{dup2} including a risk measure
\cite{cash2} indicated a crash to occur in between the end of October 1997 and
mid-November 1997. The crash occurred effectively on Monday October 27th, 1997
\cite{how} !

Eventhough the crash of October 1997 was predicted \cite{how,dup2}, the
scientific (physics or economy) \cite{critics,nonos} and media \cite{vif}
community is actually divided between those who believe in such a crash
prediction and those who believe that crashes are unpredictable events and such
findings were mere luck or at best accidental 
\cite{sornette2,feig2,brisbois}. We
discuss a little bit more the predictability problem and findings in this paper
going beyond a previous report \cite{bigtokyo}.

\subsection{Methodology and data analysis}

In e.g. Refs.\cite{bigtokyo,viz,ladek,jura},  the fact was underlined that there
are strong physical arguments stipulating that $m$ in Eq. (1) 
could be or even
should be taken as "universal". The universal $m=0$ value, i.e. a logarithmic
divergence has been proposed. The logarithmic divergence of the index 
$y$ for $t$
close to $t_c$ reads \begin{equation} y = A + B \ln{\left( {t_c-t \over t_c}
\right)} \left[ 1 + C  sin\left( \omega \ln{\left( {t_c-t \over t_c} \right)} +
\phi \right) \right] \hskip 0.3cm for  \hskip 0.3cm  t < t_c . 
\end{equation} One
should remark that the full period $[t_i,t_f]$ for a meaningful fit should
contain the whole euphoric precursor. It has been found in 
\cite{bigtokyo,how,ladek,jura} 
that the log-divergence is closer to the real signal than any
power law divergence with $m$ $\neq$ 0.

The log-divergence in Eq.(3) contains 6 parameters. At first, it seems that
non-linear fits using only the simple log-divergent function 
\begin{equation} y =
A + B \ln{(t_c-t)} \end{equation} with $B<0$, thus with only 3 
parameters can be
performed. A good estimation of $t_c$ can be obtained indeed following both
Levenberg-Marquardt and Monte-Carlo algorithms \cite{recipes}.  One 
has observed
that the estimated $t_c$ points are close to "black" days for the first two
periods \cite{bigtokyo,ladek,jura}.

Assuming that Eq.(4) is valid, one should also note that \begin{equation} {d y
\over d t} = {-B \over (t_c-t)} \end{equation}should be found in the daily
fluctuation pattern (Fig. 2). This is consistent with the volatility clustering
discussed here above. However,  Eq.(5) fits lead to bad results with huge error
bars.

The oscillating term of Eq.(3) has been quite criticized since no 
traditional or
economical argument supports the DSI theory at this time. However, the
hierarchical structure of the market has been suggested as a possible candidate
for generating DSI patterns in \cite{bigtokyo,FF96,hierarchy,FF98},so is the
price fixing ''techniques'' \cite{roehner3} and arbitrage methods. In order to
prove that a log-periodic pattern appears before crashes, the {\em envelope} of
the index $y$ is constructed \cite{viz}. Two distinct curves are 
built: the upper
envelope $y_{max}$ and the lower one $y_{min}$. The former represents 
the maximum
of $y$ in an interval $[t_i,t]$ and the latter is the minimum of $y$ in an
interval $[t,t_f]$. One observes a remarkable pattern made of a succession of
thin and huge peaks \cite{viz}.

When $y_{max}-y_{min}=0$, it means that the index $y$ reaches some value never
reached before at a time $t$ and would never have reached if the time axis had
been reversed thereafter. This corresponds to time intervals during which the
value of the index $y$ reaches new records. In fact, the pattern reflects
obviously an oscillatory precursor of the crash, thus through \begin{equation}
y_{max}-y_{min} = (C_1 + C_2 t) (1 - \cos{(\omega \ln{(t_c-t)} + \phi)})
\end{equation} where $C_1$ and $C_2$ are parameters controlling the 
amplitude of
the oscillations. The above relationship allows us to measure the log-frequency
$\omega$ \cite{viz}.  Moreover it is found that the value of $\omega$ 
seems to be
finite and almost constant, $\omega \simeq [6, 10]$ for the major analyzed
crashes. An analysis along similar lines of thought, though emphasizing the
no-divergence, thus $ m<0$ in Eqs. (1)-(2), was discussed for the Nikkei
\cite{sornettenikkei,stauijtaf} and NASDAQ April 2000 crash
\cite{sornetteNASDAQ}.

For illustrating the complexity of the frequency dependence of such financial
signals, one can also perform a Fourier transform of a reconstructed 
signal. The
evolution of the six strongest DAX crashes between Oct. 01, 1959 and Dec. 31,
1988 and the strongest DAX crashes in 1997, prior to the crash day, 
are shown in
Fig. 4; $y_0$ denotes the index value at the closing of the crash day. Power
spectra of the DAX index measured from the index value at the end of the crash
day have been calculated for a time interval equal to 600 days, those prior to
crashes. The 6 large DAX crash spectra for the period of interest are shown in
Fig. 5. The corresponding exponents $\beta$ for the best fit in the high
frequency region are given in Table 1.

\begin{figure}
\includegraphics[width=.9\textwidth]{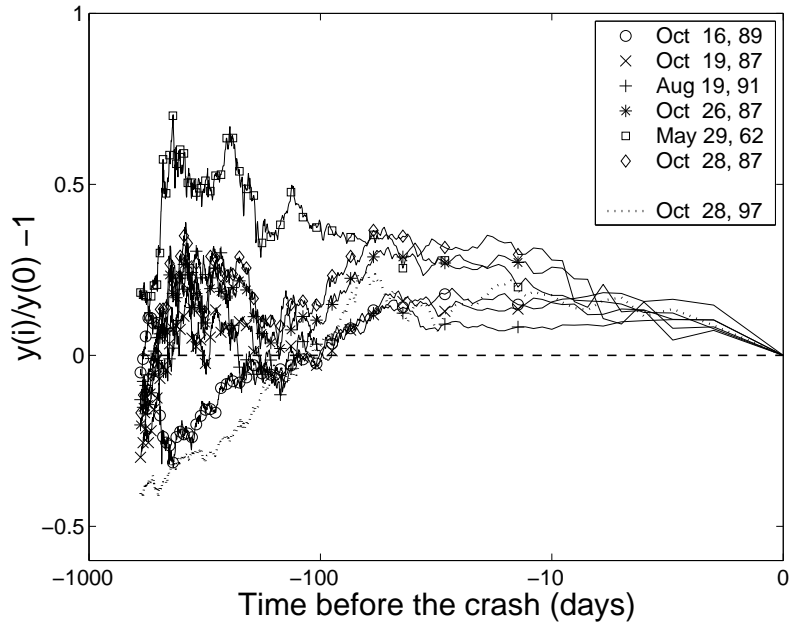}
\caption[]{The evolution of
the six strongest DAX crashes between Oct. 01, 59 and Dec. 31, 88, and the
strongest DAX crash in 1997, prior to the crash day ; $y_0$ denotes the index
value at the closing of the crash day}
\label{eps4}
\end{figure}

\begin{figure} \centering 
\includegraphics[width=.48\textwidth]{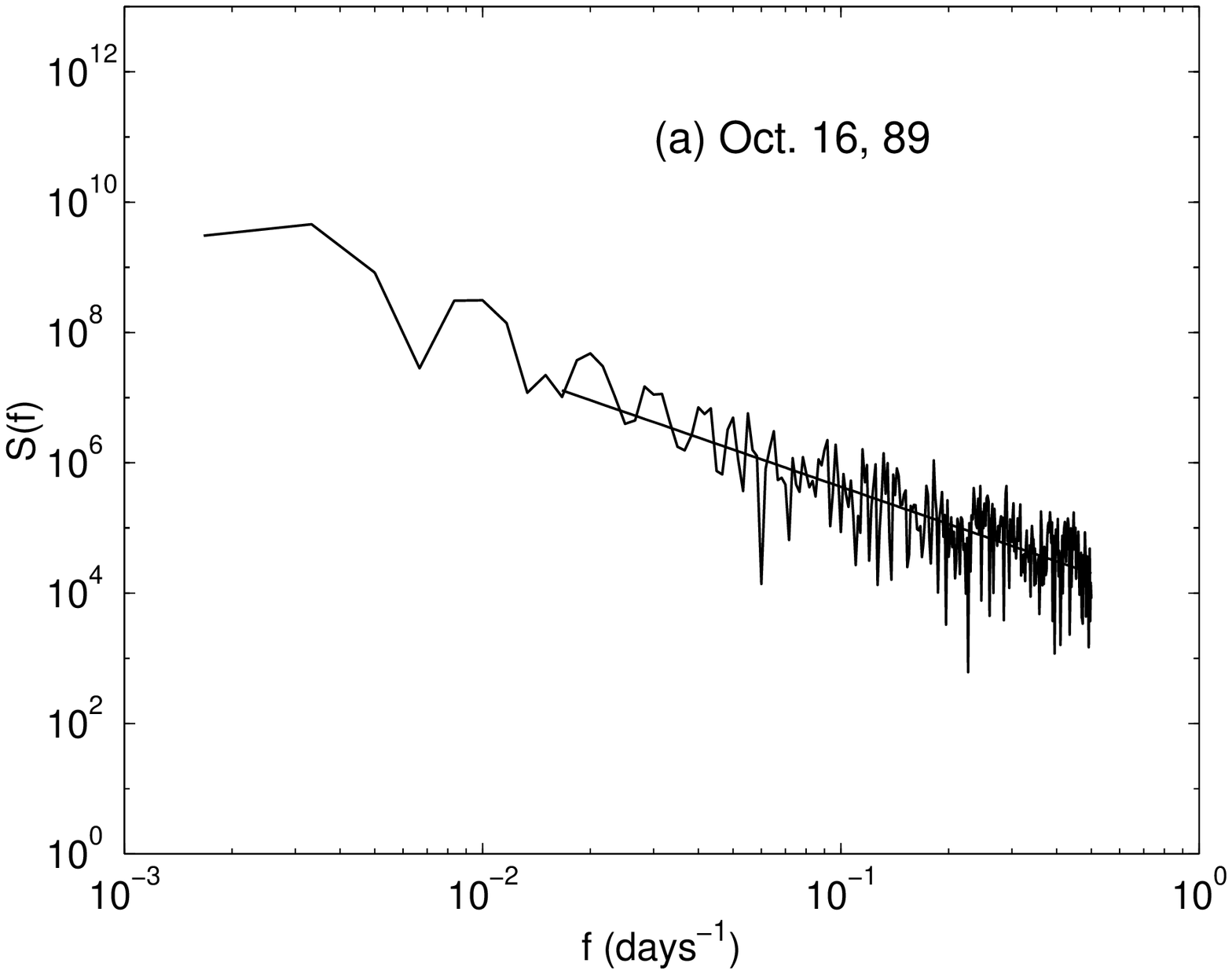} \hfill
\includegraphics[width=.48\textwidth]{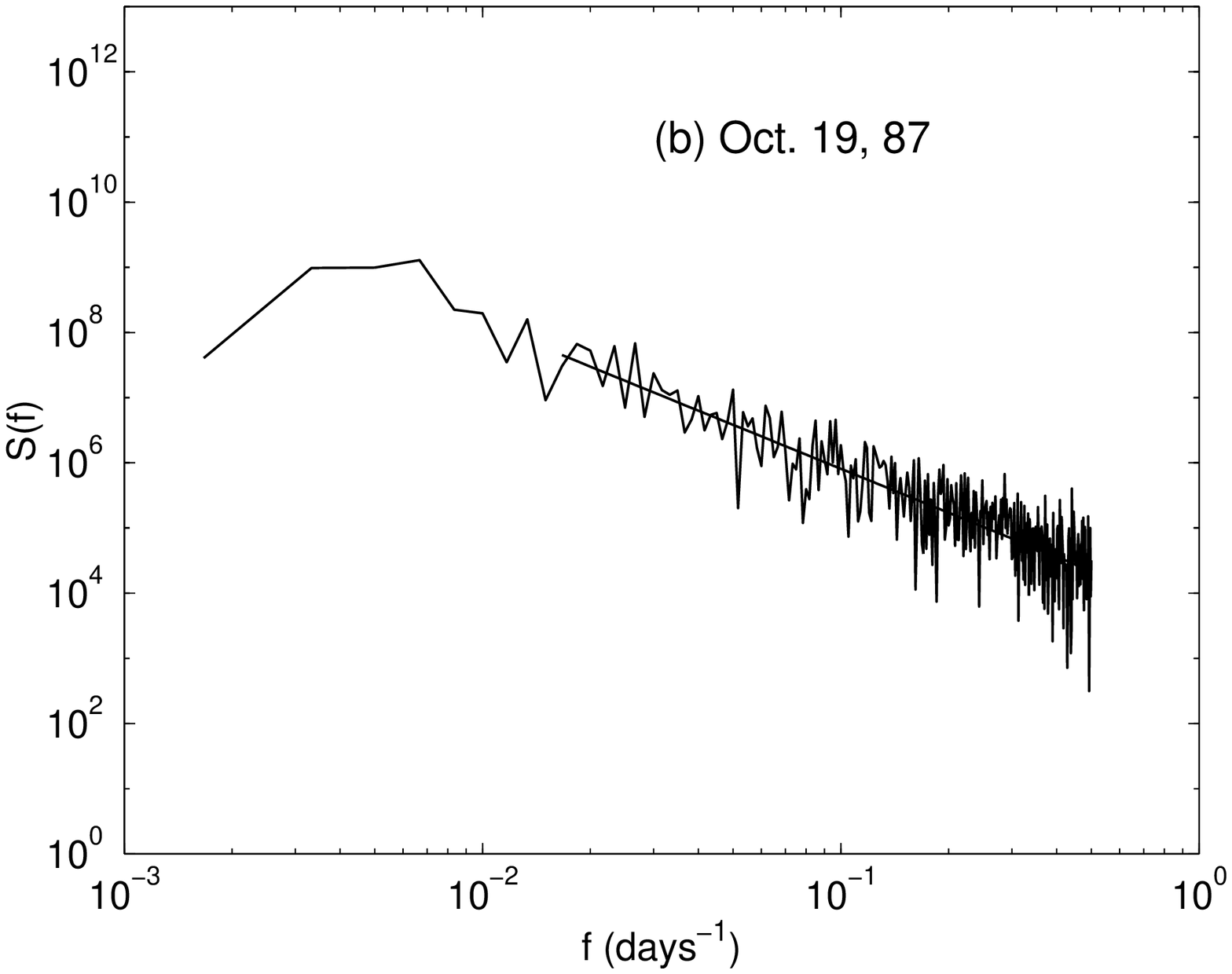} \vfill
\includegraphics[width=.48\textwidth]{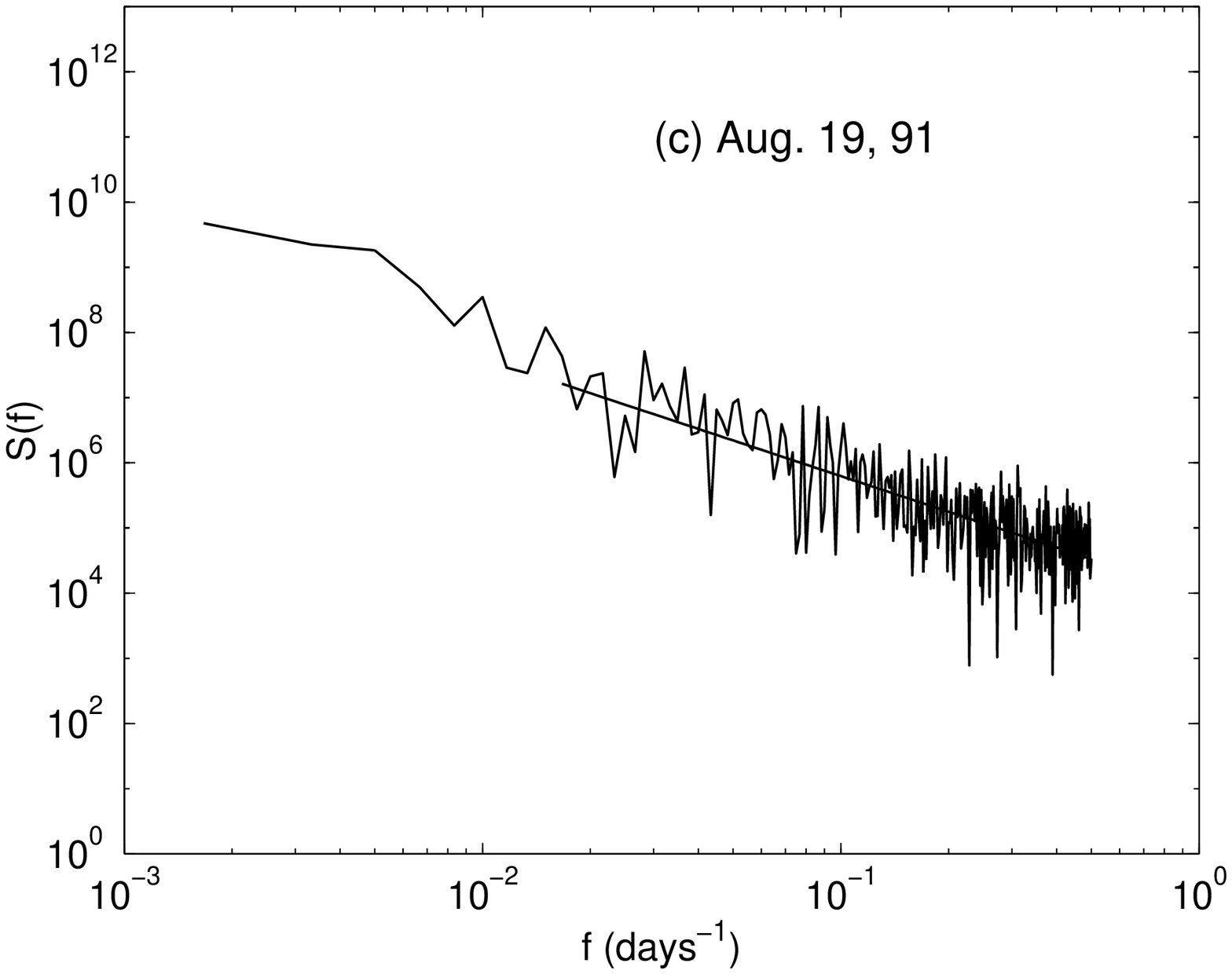} \hfill
\includegraphics[width=.48\textwidth]{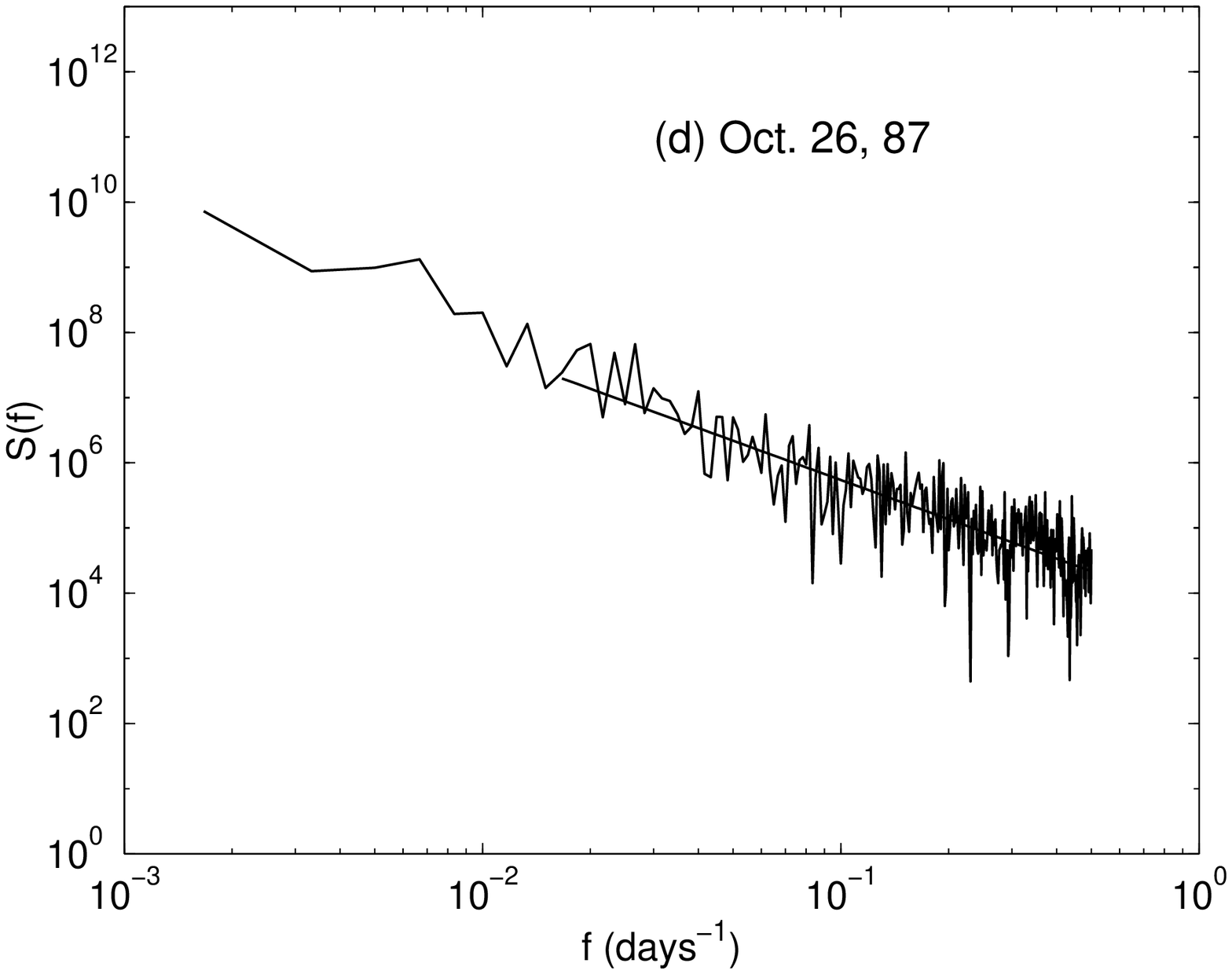} \vfill
\includegraphics[width=.48\textwidth]{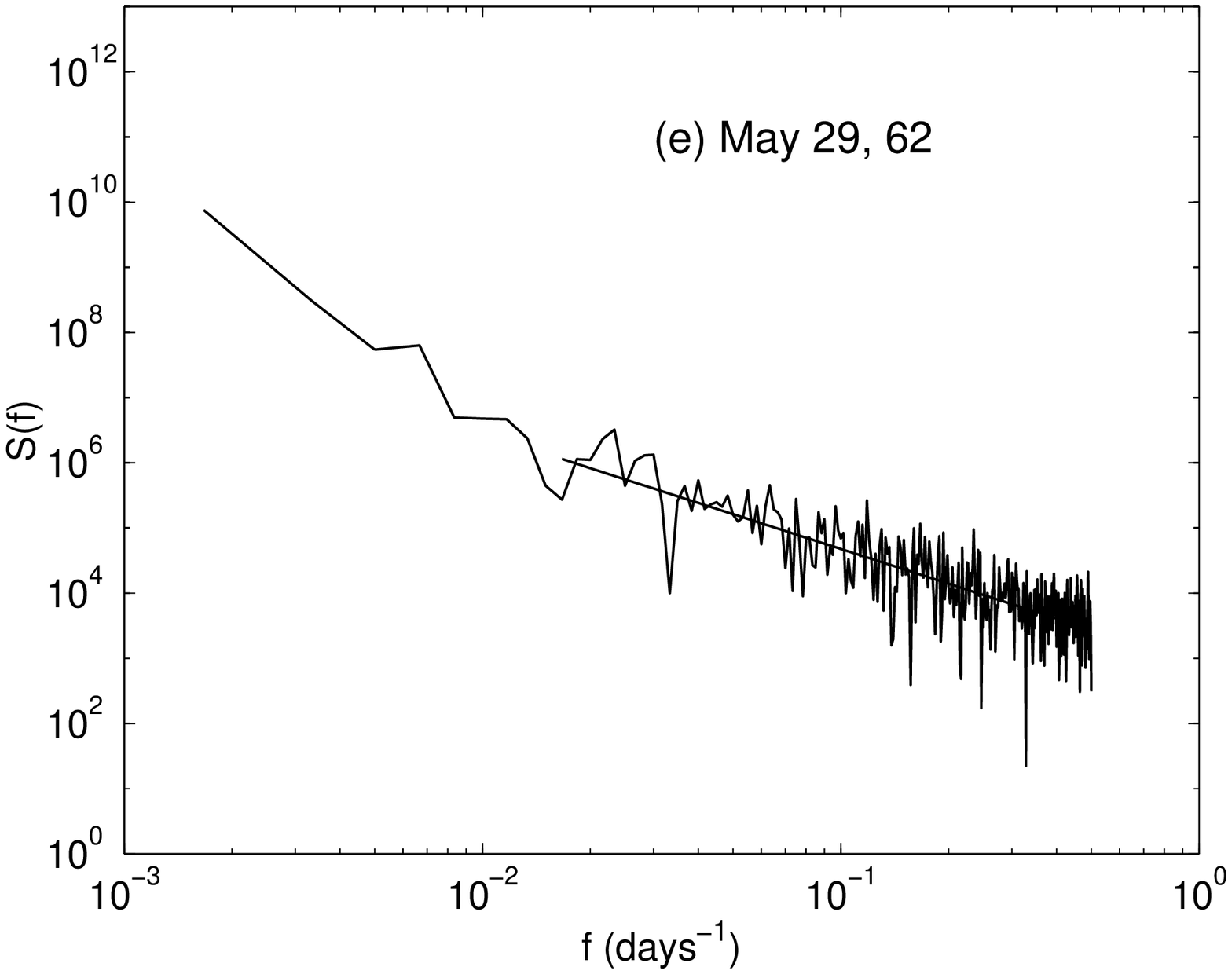} \hfill
\includegraphics[width=.48\textwidth]{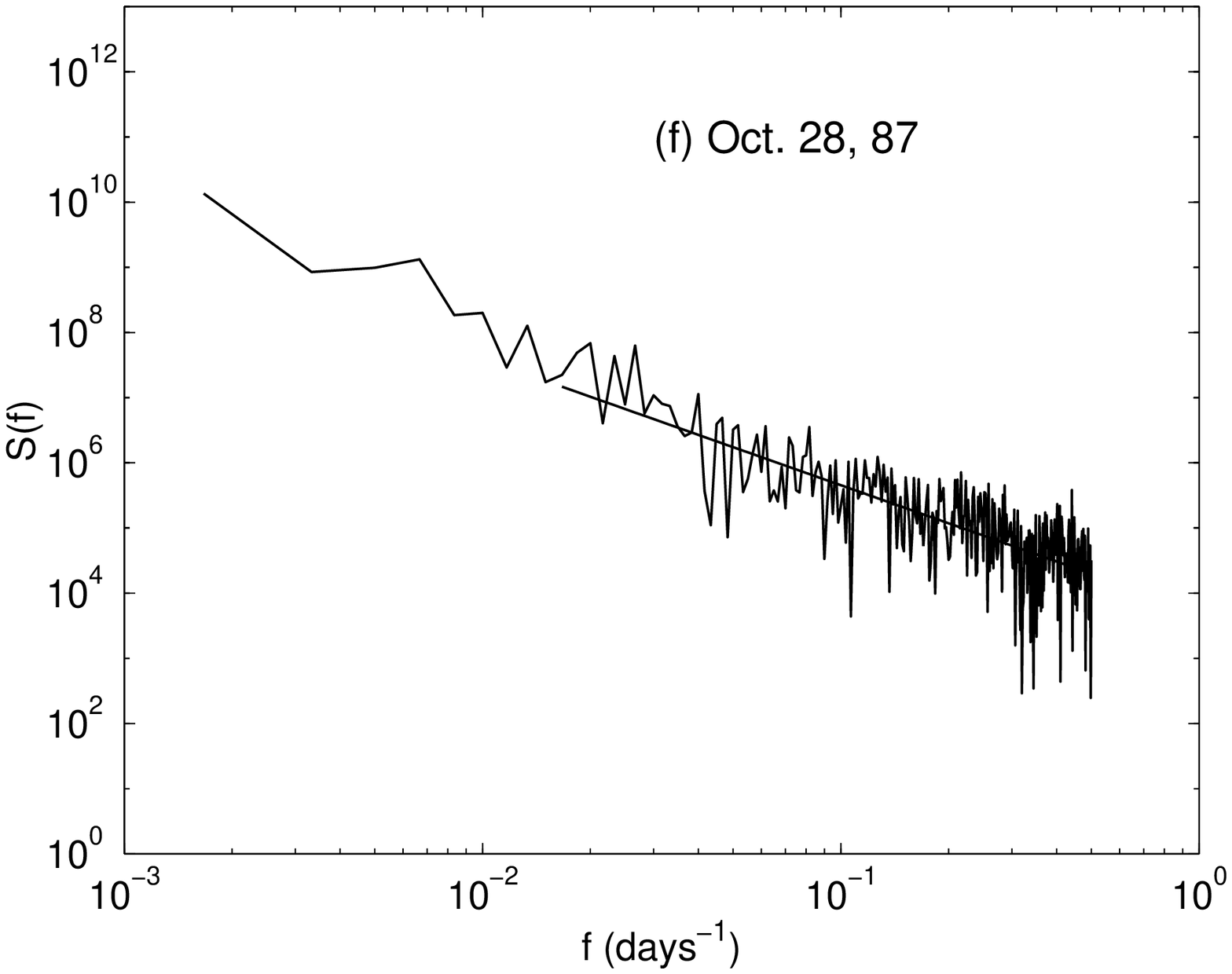} \vfill
\caption[]{The power
spectrum of the 600 day DAX index evolution signal corresponding to 
the six major
crashes between Oct. 01, 59 and Dec. 31, 96}
\label{eps5}
\end{figure}

The roughness behavior \cite{schroeder,west,chemnitz} of the DAX 
index evolution
signal before crashes can be defined trough the fractal dimension $D$ of the
signal, i.e. \cite{schroeder} \begin{equation} D = E + \frac{3 - \beta}{2},
\end{equation} where $E$ is the Euclidian dimension. The values of $\beta$ and
$D$ are reported in Table 1 with the crash dates and relative 
amplitude of the 6
major DAX crashes which occurred between Oct. 01, 1959 and Dec. 30, 1996. The
same type of data is reported in Table 2 for the 3 major DAX crashes 
in  October
1997. The power spectrum of the large Oct. 28, 97 crash is shown in Fig. 6.

\begin{figure}
\includegraphics[width=.9\textwidth]{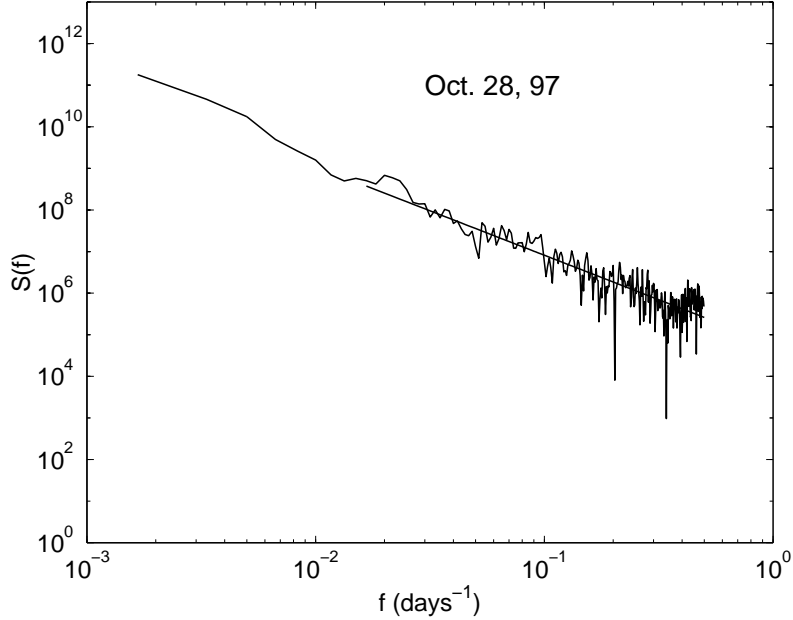}
\caption[]{The power
spectrum of the 600 day DAX index evolution signal prior to the Oct. 
28, 97 crash}
\label{eps6}
\end{figure}

\begin{table}[ht] \centering \caption{Crash dates and relative 
amplitude of the 6
major DAX crashes having occurred between Oct. 01, 1959 and Dec. 30, 
1998; power
law exponent $\beta_-$ of the signal power spectrum  (600 data points) and
corresponding fractal dimension $D_-$ of the signal prior to, and $\beta_+$ and
$D_+$, after these 6 major DAX crashes.} \renewcommand{\arraystretch}{1.4}
\setlength\tabcolsep{5pt} \begin{tabular}{cccccc} \hline  crash dates 
& relative
amplitude & $ \beta_-$ & $ D_- $ & $ \beta_+$ & $ D_+ $\\ \hline 
16.10.89& -0.137
&1.90$\pm$0.09&1.55&1.67$\pm$0.09&1.67\\ 19.10.87& -0.099
&2.24$\pm$0.09&1.38&1.82$\pm$0.09&1.59\\ 19.08.91& -0.099
&1.82$\pm$0.10&1.59&1.62$\pm$0.06&1.69\\ 26.10.87& -0.080
&2.00$\pm$0.09&1.50&1.70$\pm$0.08&1.65\\ 29.05.62& -0.075
&1.77$\pm$0.08&1.62&1.28$\pm$0.08&1.86\\ 28.10.87& -0.070
&1.94$\pm$0.10&1.53&1.81$\pm$0.07&1.60\\

\hline \end{tabular} \label{Tab1} \end{table}

\begin{table}[ht] \centering \caption{The 3 major DAX crashes in October 1997 :
crash dates, relative amplitude, power law exponent $\beta_-$ of the 
signal power
spectrum  (600 data points prior to these 3 major DAX crash dates) and the
corresponding fractal dimension $D_-$.} \renewcommand{\arraystretch}{1.4}
\setlength\tabcolsep{5pt} \begin{tabular}{cccc} \hline  crash dates  & relative
amplitude & $ \beta_- $ & $ D_- $ \\ \hline

28.10.97& -0.084 &2.14$\pm$0.07&1.43\\ 27.10.97& -0.043 &1.97$\pm$0.07&1.52\\
23.10.97& -0.048 &1.91$\pm$0.05&1.55\\

\hline \end{tabular} \label{Tab2} \end{table}

\subsection{Aftershock patterns}

The index evolution after a crash has also been analyzed through a 
reconstructed
signal that is the difference between the DAX value signal at each 
day $y(i)$ and
the DAX value at the crash day $y_0$. For the 6 largest crashes in the time
interval of interest the recovery can be slow (Fig.7). It took about one month
for the Oct. 28, 97 crash. To observe some periodic fluctuation after 
the crash,
the power spectrum of the DAX has been computed for the 600 days following a
crash day (Fig. 8 (a-f)). Note the high-frequency log-periodic 
oscillation regime
of the power spectrum for the Oct. 19, 1987 case on Fig. 7(d).   The values of
each $\beta$ and corresponding fractal dimension $D$ are reported in Table 1.

\begin{figure}
\includegraphics[width=.9\textwidth]{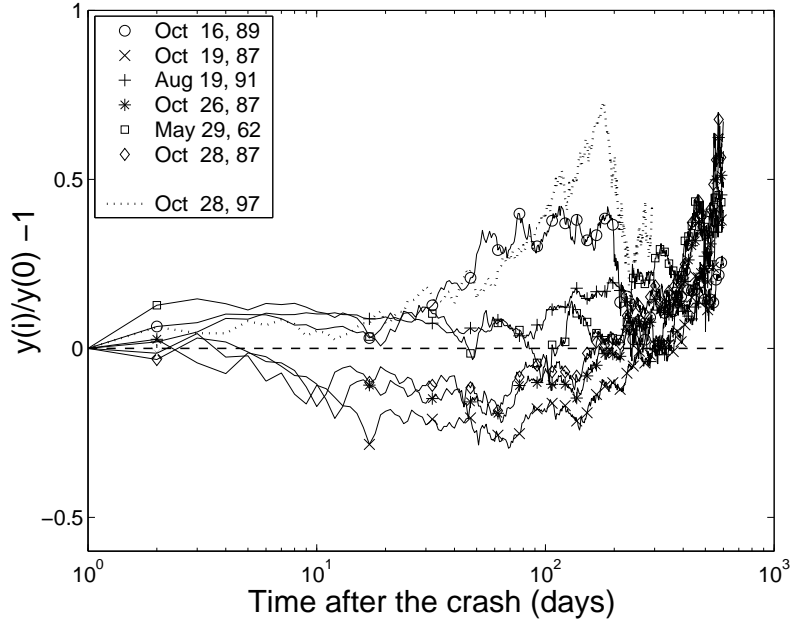}
\caption[]{The DAX recovery
signal evolution after the six strongest DAX crashes having occured 
between Jan.
01, 85 and Dec. 31, 88, and after the strongest DAX crash in 1997 ; 
$y_0$ denotes
the index value at the closing of the crash day}
\label{eps7}
\end{figure}

\begin{figure} \centering 
\includegraphics[width=.48\textwidth]{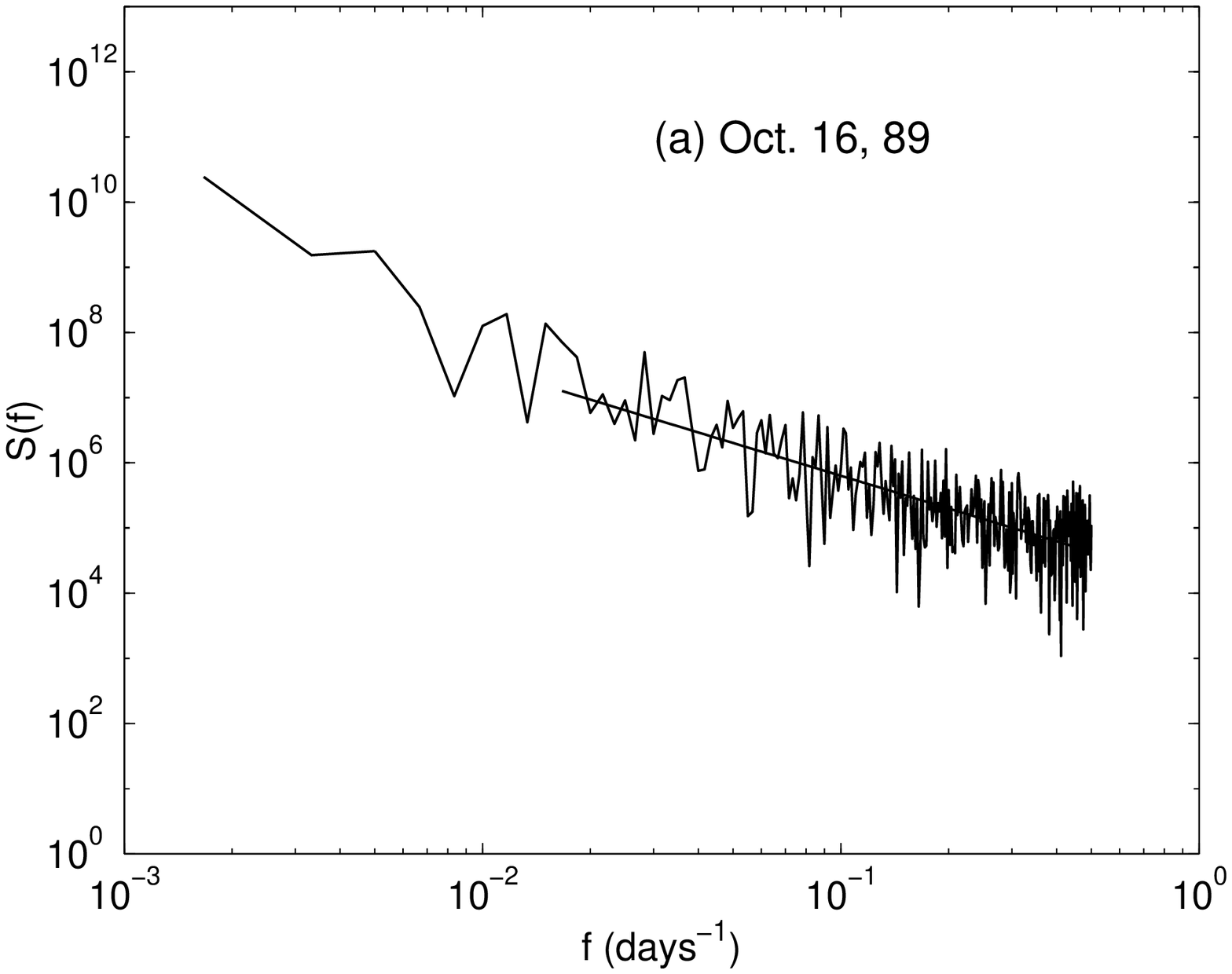} \hfill
\includegraphics[width=.48\textwidth]{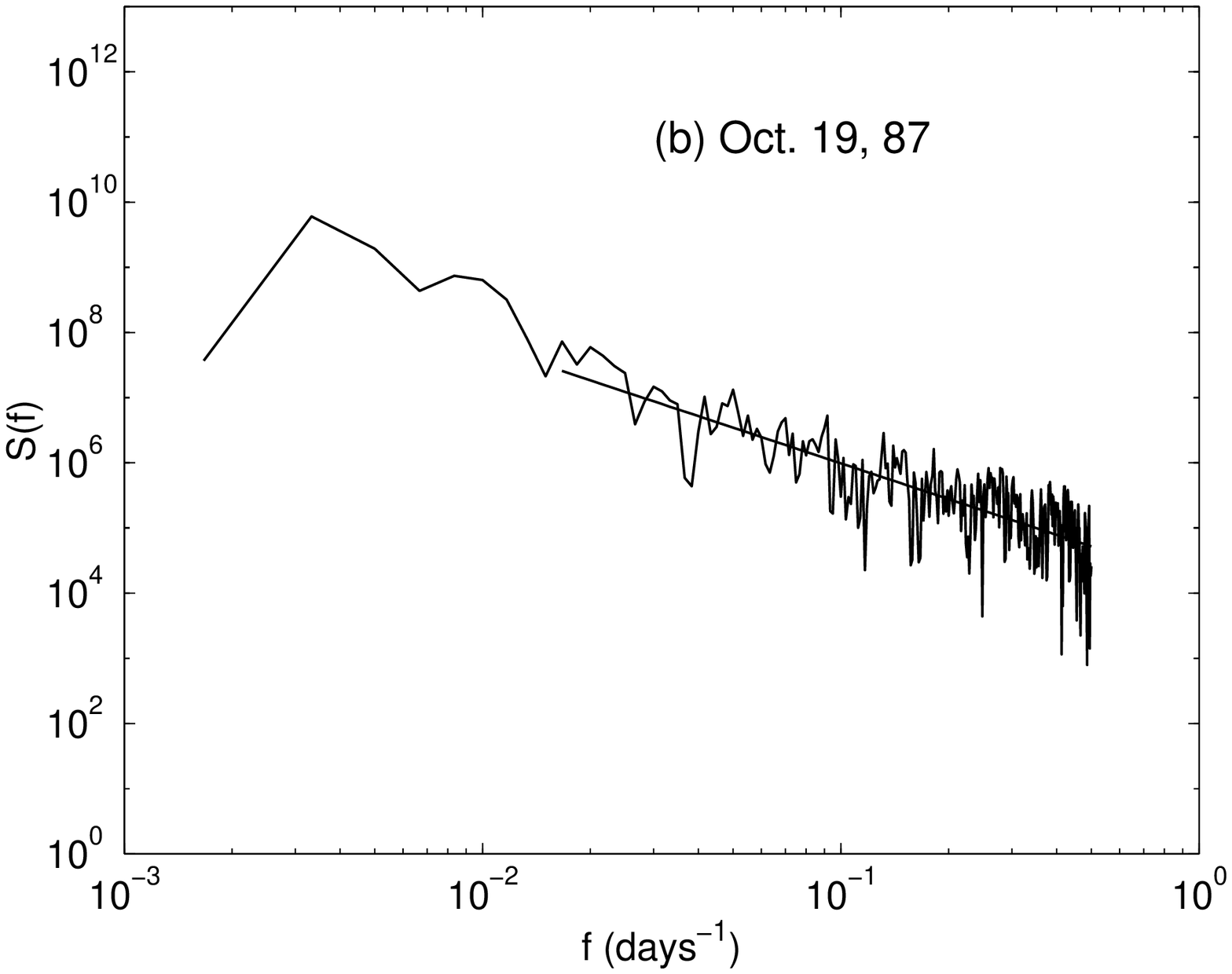} \vfill
\includegraphics[width=.48\textwidth]{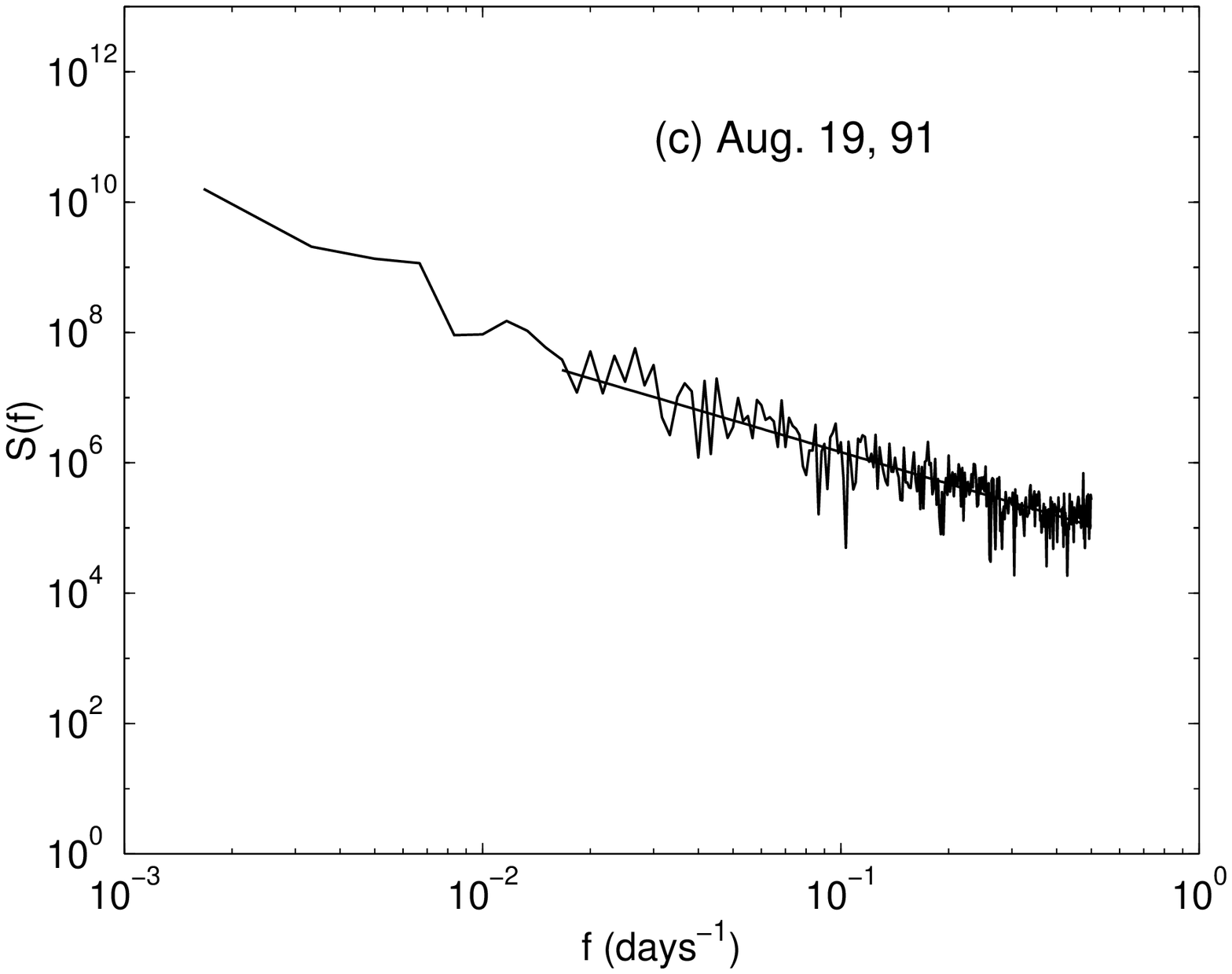} \hfill
\includegraphics[width=.48\textwidth]{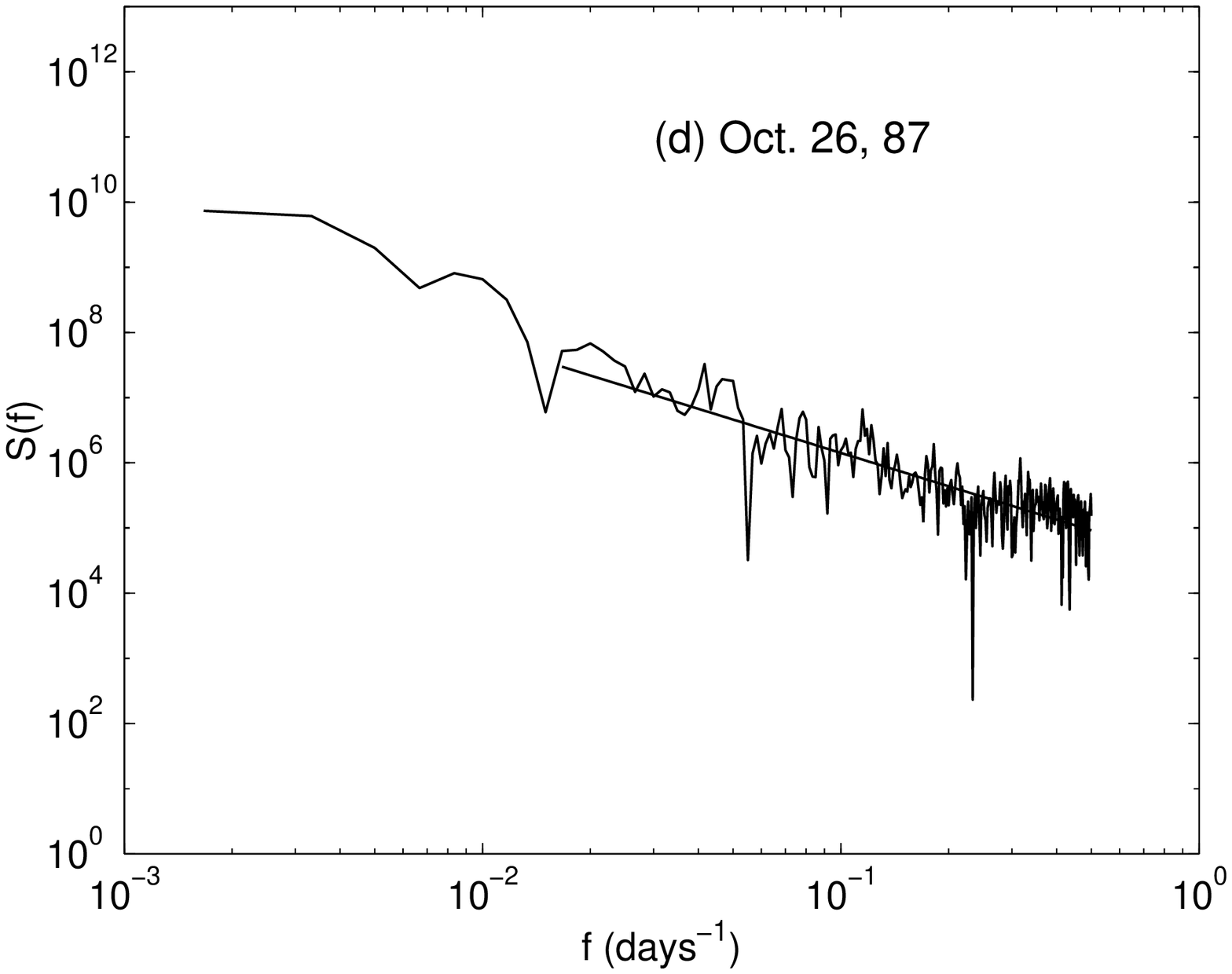} \vfill
\includegraphics[width=.48\textwidth]{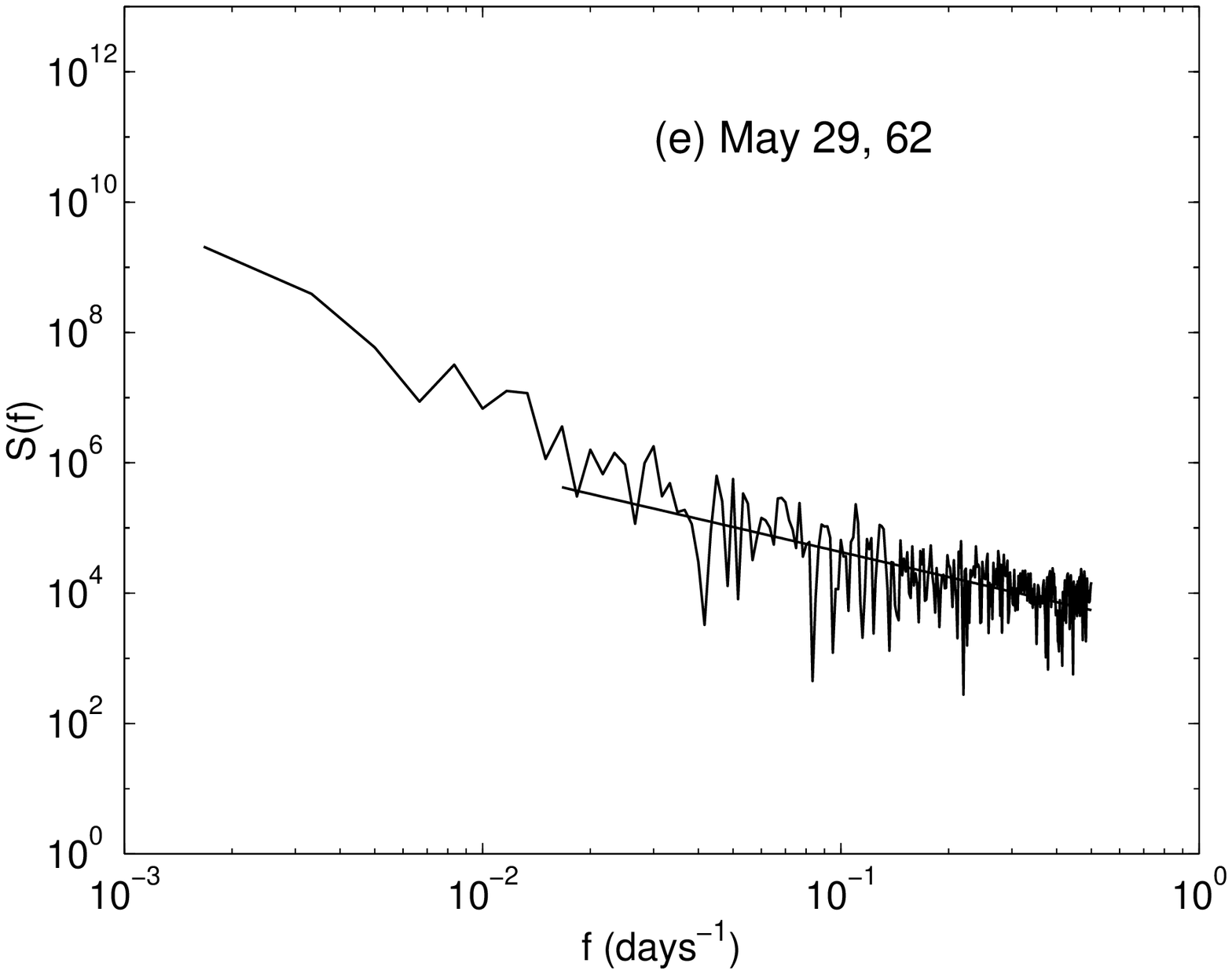} \hfill
\includegraphics[width=.48\textwidth]{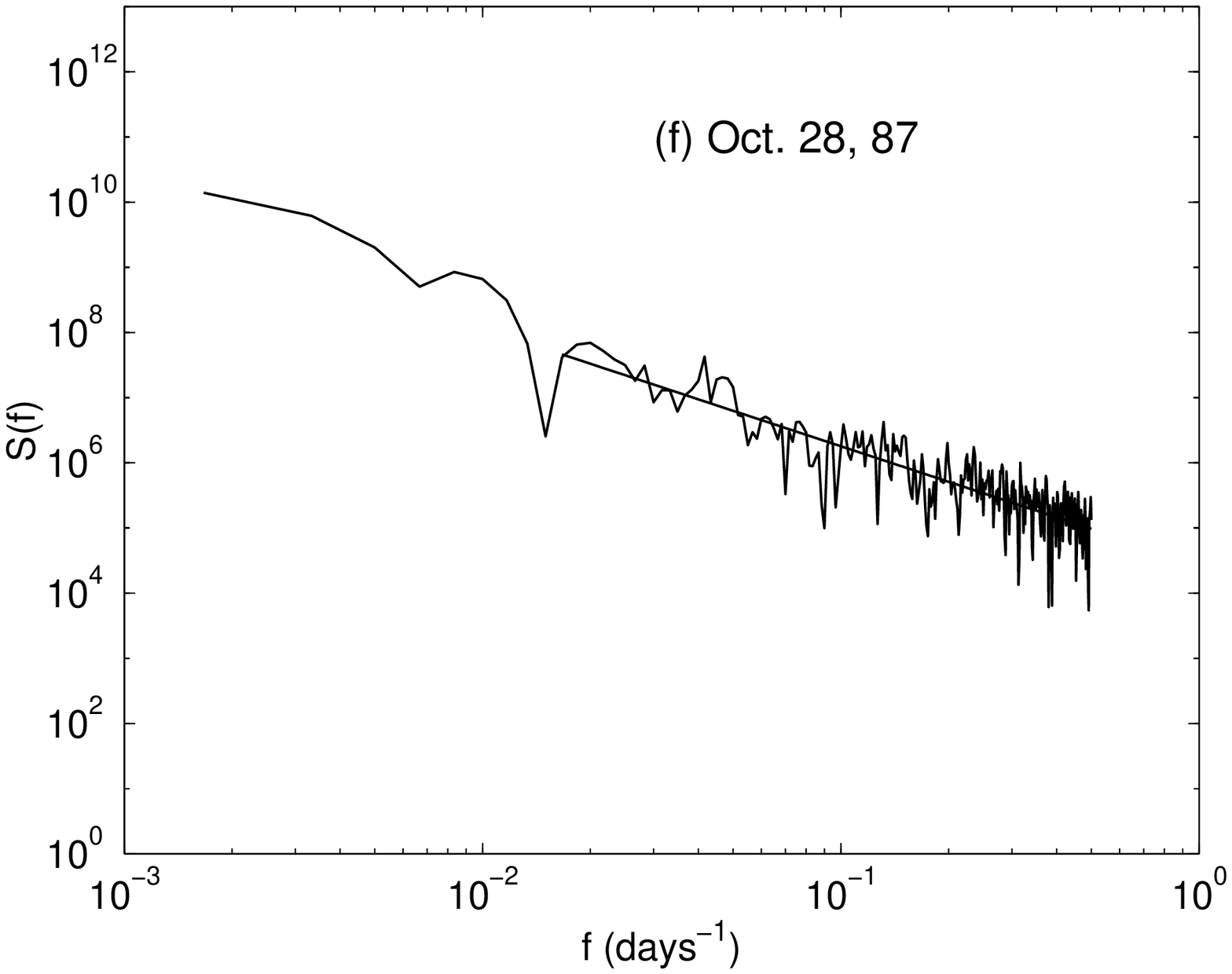} \vfill
\caption{The power
spectrum of the 600 day DAX index evolution signal after the six major crashes
having occured between Oct. 01, 59 and Dec. 31, 96} 
\label{eps8} 
\end{figure}

As a final point of this section, it should be noticed that the 
fractal dimension
is close to 1.70, thus {\it very similar} to that of a percolation 
backbone. This
might be the hint that hierarchical structures are present, and a cause of
crashes. As a consequence, the market could be viewed as a discrete fractal
system, transiting at crashes like a physical system at a percolation
transition..  In related work, Amaral and coworkers \cite{amaral} have studied
the statistics of several companies as well as their respective 
growth. They have
found that the growth of companies can be modelled using a hierarchical lattice
like a Cayley tree. For simple models of hierarchically organized markets  some
self-regulation is found in fact \cite{sopthierarchy}. On such systems the
fractal dimension can be considered to have an imaginary part which 
is related to
the log-periodic oscillations, - in fact is the signature of the 
branching ratio
\cite{fractaltree}.

In conclusion of this section,  we may conjecture that stock markets are also
hierarchical objects where each level may have a different weight, connectivity,
and characteristics time scale (the horizons of the inve 
stors) \cite{bigtokyo}.
The hierarchical tree might be fractal at crashes and its geometry 
might control
the type of criticality. This gives some argument in favor of the 
sand pile model
on a fractal basis \cite{fractalsand} as a microscopic model actually able to
simulate a crash \cite{bigtokyo}.

\section{Foreign Currency Exchange Rates}

Beside the crash cases discussed here above numerous examples of 
scale invariance
seem to be widespread in natural and social systems \cite{west,bakbook}. A
fundamental problem is the existence and width of the scaling range for
long-range power-law correlations (LRPLC) in economic systems, as well as the
presence of economic cycles. Indeed, traditional methods (like 
spectral methods)
have corroborated the evidence that the Brownian motion idea or ordinary random
walk is quite away from reality and LRPLC quite frequent
\cite{stanley,peters1,MantegnaStanleybook,voit}. Different approaches
\cite{chemnitz} have been envisaged to measure the LRPLC or analyze them in
financial data: tails of partial distribution functions of the volatility,
wavelet analysis, Detrended Fluctuation Analysis (DFA) \cite{DNADFA}, etc.

\subsection{DFA analysis} \vskip 0.6cm

The DFA method \cite{DNADFA} consists in dividing the whole data 
sequence $y(n)$
of length $N$ into $N/\tau$ non overlapping boxes, each containing 
$\tau$ points.
Then, the {\it local trend} \begin{equation} z(n)=an+b \end{equation} 
in each box
is defined to be the ordinate of a linear least-square fit of the 
data points in
that box. One should remark that a trend $z(n)$ different from a first-degree
polynomial can also be used like the cubic trend \cite{ndub}. Other detrending
functions may improve the accuracy of the DFA technique, sort out the 
reason for
crossovers between scaling regimes, and pin point noise and intrinsic trends
\cite{hu}.

The so-defined detrended fluctuation function $F(\tau)$ is then calculated
following \begin{equation} {{F(\tau)}^2} = {1 \over \tau}
{\sum_{n=kt+1}^{(k+1)\tau} {|y(n)-z(n)|}^2} , {\hskip 1cm} k=0,1,2, \cdots,
\left( {N \over \tau} -1 \right) \end{equation} Averaging $F(\tau)$ over the
$N/\tau$ intervals gives a function depending on the box size $\tau$. The above
calculation is repeated for different box sizes $\tau$. If the $y(n)$ data are
randomly uncorrelated variables or short range correlated variables, 
the behavior
is expected to be a power law \begin{equation} \langle F(\tau)^2 \rangle^{1/2}
\sim \tau^{\alpha} \end{equation} with an exponent 1/2 \cite{DNADFA} if the
excursion is governed by a mere random walk. An exponent $\alpha 
\not= 1/2$ in a
certain range of $\tau$ values implies the existence of LRPLC in that time
interval. Mathematically, the correlation of a future increment 
$y(n)-y(0)$ with
a past increment $y(0)-y(-n)$ is given by \begin{equation} \Gamma (n) 
= {\langle
(y(0)-y(-n))(y(n)-y(0)) \rangle \over \langle (y(n)-y(0))^2 \rangle} = 2^{2
\alpha -1}-1, \end{equation} where the correlations are normalized by the
variance of $y(n)$. For $\alpha > 1/2$, there is {\em persistence}, i.e.
$\Gamma>0$. In this case, if in the immediate past the signal has a positive
increment, then on the average an increase of the signal in the 
immediate future
is expected. In other words, persistent stochastic processes exhibit 
rather clear
trends with relatively little noise. An exponent $\alpha < 1/2$ means {\em
antipersistence}, i.e. $\Gamma<0$. In this case, an increasing value in the
immediate past implies a decreasing signal in the immediate future, while a
decreasing signal in the immediate past makes an increasing signal in 
the future
probable. In so doing, data records with $\alpha < 1/2$ appear very {\it noisy}
(rough). The $\alpha = 0$ situation corresponds to the so-called {\it white
noise}. Finally, one should note that $\alpha$ is nothing else than $Ha$,  the
so-called Hausdorff exponent for fractional Brownian motions
\cite{west,chemnitz}. It can be useful to recall \cite{chemnitz} that the power
spectrum of such random signals is characterized by a power law with 
an exponent
$\beta = 2 \alpha - 1$.

\subsection{Data and analysis}

We have considered the daily evolution of several currency exchange rates with
respect to the $USD$ from January 1990 till December 1999 including 
only all open
banking days. This represents about $N=3000 $ data points.  The data are those
obtained from \cite{quoteEUR}, at the closing time of the foreign 
exchange market
in London for the ten currencies $C_i$, (i=1,10) forming the $EUR$ on Jan. 01,
1999.

\begin{figure}
\includegraphics[width=.9\textwidth]{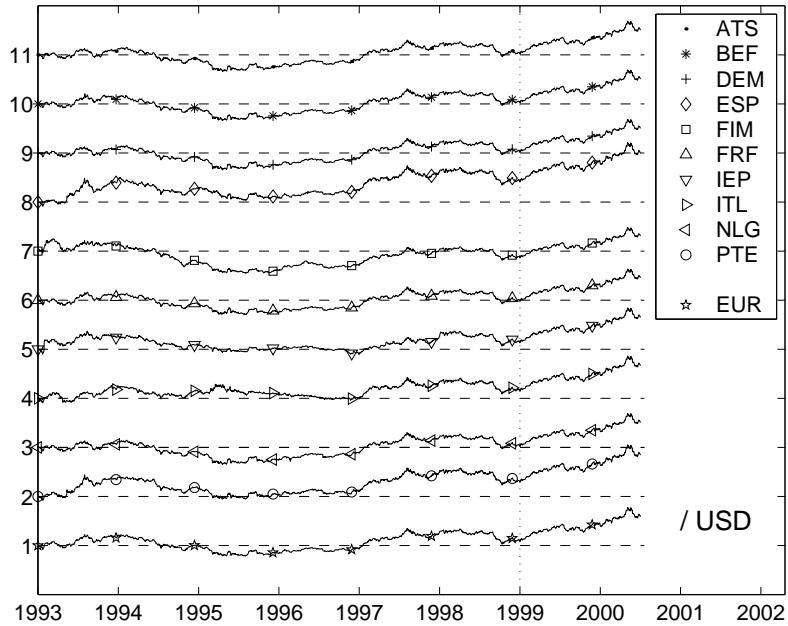}
\caption[]{Normalized $EUR$ and $C_i$ (i=1,10) 
currency forming
the $EUR$ exchange rates with respect to the $USD$ between Jan. 01, 
1993 and June
30, 2000. The data are artificially multiplied by two and then displaced along
the vertical axis in order to make the fluctuations noticeable. The 
vertical dash
line marks the date for the EUR official introduction}
\label{eps9}
\end{figure}

The evolution of such $C_i$/$USD$ exchange rate from Jan. 01, 1993 to June 30,
2000 is drawn in Fig. 9. In Fig. 10, a log-log plot of the 10 
functions $\langle
F(t)^2 \rangle^{1/2}$ is shown for the whole data of Fig. 9. Moreover 
we plot the
result for a false $EUR$, i.e. a linear combination of the ten 
currencies forming
the $EUR$ \cite{Ref1EUR,kimalg,tokyokima,ijmpc} Except for $IEP$, the functions
are very close to a power law with an exponent $\alpha = 0.51 \pm 0.02$ holding
over two decades in time, i.e. from about one week to two years. This finding
clearly shows the non-existence of LRPLC in the foreign exchange market with
respect to the $USD$. Other cases showing marked deviations from 
Brownian motion
have been discussed elsewhere \cite{kimalg,ijmpc,nvma,h1c1}.  It can then be
observed that a wide variety of behaviors is found in the foreign currency
exchange market. Exponent values and the range over which a power law holds
drastically  vary from a currency exchange rate to another. It appears that the
currency exchange rates can be classified into three different categories from
the LRPLC point of view.

\begin{figure}
\includegraphics[width=.9\textwidth]{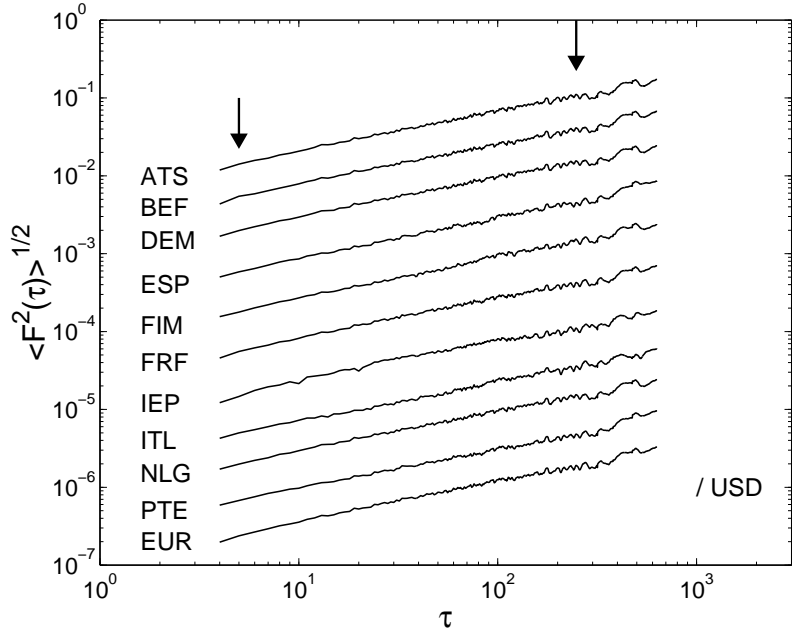}
\caption[]{Log-log plot of the DFA function showing how to
obtain the $\alpha$ exponent for the 11 exchange rates of interest 
for $EUR/USD$.
The fit slope being only of interest, the DFA function data has been 
arbitrarily
displaced along the vertical axis}
\label{eps10}
\end{figure}

First, the rates which exhibit an exponent $\alpha$ larger than 1/2 (persistent
behavior). This case corresponds to currency exchange rates between leading
currencies (e.g., $USD$, $JPY$, $EUR$) and so called $weaker$ ones
\cite{h1c1,kiagina}.

A second category concerns the rates exhibiting strict randomness 
($\alpha=1/2$)
within error bars. This is the case for example of the $USD/{C_i}$ 
rates as shown
above.

A third category represents the currency exchange rates with antipersistent
behavior ($\alpha<1/2$) as e.g. $DEM/BEF$ \cite{nvma}. These currencies most
often concern currency exchange rates between  (european) countries which are
submitted to strict monetary rules and to strict regulatory corrections by
central banks due to international multilateral conventions. It 
should be pointed
out that in general the range, over which the antipersistency 
signature, i.e. the
power law is valid, occurs over a limited time span in this third category. In
fact, there is a crossover around $\tau^* \approx 10$ weeks. For longer time
scales ($\tau >> \tau^*$), the signal becomes again persistent or random.

\subsection{Probing the local correlations}

It is also of interest to know whether the LRPLC are stable along the data. In
order to  probe the local strength of the correlations, one constructs a
so-called observation box of width $T$ placed at the beginning of the data, and
calculates $\alpha$ for the data contained in that box. Then, the box is moved
along the data by some step  toward the right along the financial sequence and
$\alpha$ is again calculated. Iterating this procedure one obtains a "local
measurement" of the degree of "long-range correlations" over $T$. It is crucial
to choose the most adequate box size $T$, i.e. to choose $T$ of the 
same order of
magnitude as the maximum range $\tau$ over which the above power law is valid.

\begin{figure}
\includegraphics[width=.9\textwidth]{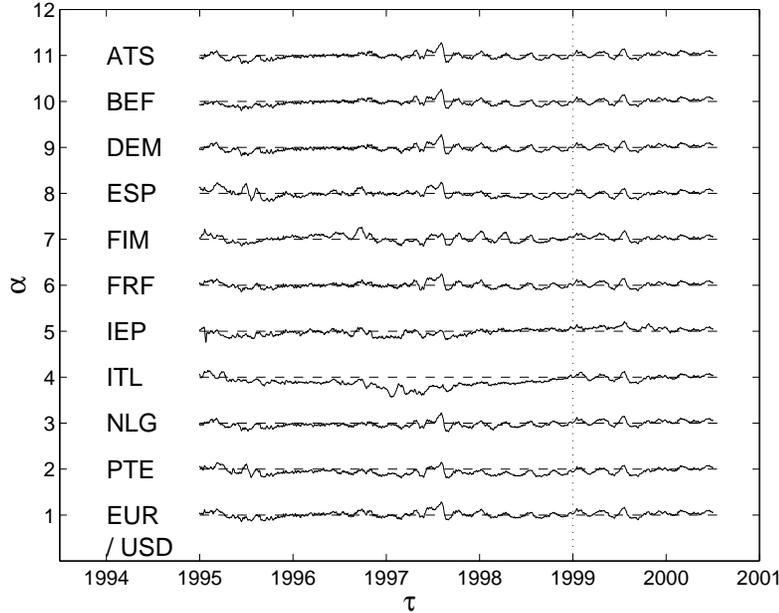}
\caption[]{Time dependence of the DFA $local$
$\alpha$-exponent for $EUR$ and each currency forming the $EUR$ exchange rate
with respect to  $USD$. The $\alpha$-values are artificially multiplied by two
and then displaced along the vertical axis in order to make the fluctuations
noticeable. For each time dependent $\alpha$ a horizontal dashed line 
is drawn to
indicate a reference to Brownian fluctuations}
\label{eps11}
\end{figure}

The evolution of the $EUR/USD$ and $C_i/USD$ $\alpha$'s for the 
1995-2000 period
is illustrated in Fig. 11.  In order to probe the local values of $\alpha$, we
have used a window of size $T=2$ years. The exponent $\alpha$ varies 
around 1/2,
i.e. the horizontal dashed line in Fig.10. The local value of $\alpha$ seems to
decrease at first and regrows in 95, is stable in 96,   has a big 
fluctuation in
mid/fall 97 and becomes pretty stable thereafter. The ITL case evolution is
slightly different. The minute differences have probably to be associated to
national political or economic  events having an impact on the international
monetary policy. It seems interesting to notice that the large fluctuations in
$\alpha$ occur just before the crash dates of stock market indices. 
See also the
marked singularity in mid 1999,  a signature of the XR's adjustments 
prior to the
EUR  introduction. In order to further {\it prove} this point, a linear DFA
analysis of the Dow Jones Industrial stock index around the 1987 October crash
was performed \cite{nvmaunpublished}. A similar pattern is found.

Other XR time series have been examined in order to check the 
non-stationarity of
$\alpha$ \cite{kimalg,ijmpc}. This does support the idea that the foreign
currency exchange markets are mainly governed by random conditions
\cite{friedrich} or is said to be {\it efficient} in more usual economic
language. However, this unconditional randomness cannot be extrapolated to
speculating times nor emerging currencies \cite{kilev}. Different universality
classes thereby emerge. It may be useful to recall that Hartmann\cite{hartmann}
has examined the competition between $USD$ and $EUR$ in a more 
general (political
and economy) framework.

\section {Conclusions}

The DAX has been analyzed from the point of view of crashes, in particular the
correlations in the signal volatility, before and after the critical days. The
search for the crash day is separated into two numerical problems, that of the
index divergence itself and that of the index oscillation frequency 
acceleration
on the other hand.  By considering the envelope of the DAX, we have 
demonstrated
that before crashes, a log-periodic pattern exists. Even though error bars are
intrinsically large, it is surprising to see that a rupture point is easily
predicted. A hierarchical structure close to a fractal percolation backbone (or
tree) seems intrinsic at crashes. The stability of this result should be tested
in real time for the best future of our economic system. A few foreign exchange
currency rates with respect to the $USD$ have been examined in order to
illustrate the DFA technique, the intrinsic structure of the DFA exponent, and
its implications with respect to crashes.

\vskip 1.0cm {\noindent \large Acknowledgements} \vskip 0.6cm

Luc T. Wille is gladly thanked for inviting us to present the above results and
considerations, and enticing us into writing this report. Thanks to 
the State of
Florida for some financial support allowing the authors to participate in the
conference.  MA thanks N. Vandewalle for numerous discussions.

\end{document}